\magnification=\magstep1 
\hsize=16.5truecm
\vsize=24truecm
\hfuzz=2pt 
\vfuzz=4pt
\pretolerance=500
\tolerance=1000
\parskip=0pt
 \parindent=8.5truemm

\font\fourteenbf=cmbx10 scaled 1360        

{\vglue 5truecm{
               \baselineskip=24 truept
               
               \raggedright
               \parskip=0pt
               \noindent
               \fourteenbf  {REPRESENTATIONS OF CLASSICAL GROUPS 
               ON THE LATTICE AND ITS APPLICATION TO THE FIELD THEORY  ON DISCRETE SPACE-TIME}
\vskip 1.3truecm}

{
 \parindent=25.4mm  Miguel Lorente \par 
 \medskip Departamento de F{\'\i}sica \par
Universidad de Oviedo \par 
33007 Oviedo, Spain }

\bigskip {\noindent We explore the mathematical consequences of the assumption of a discrete
space-time. The fundamental laws of physics have to be translated into the language of discrete
mathematics. We find integral transformations that leave the lattice of any dimension invariant and
apply these transformations to field equations.} 

\vskip 10mm {\noindent \bf 1. INTRODUCTION}

\medskip The idea of a discrete space-time has been introduced by physicists in the past in
several different ways. Heisenberg advocated a fundamental length, Snyder has proposed a
position operator with discrete spectrum, Ahmavaara has carried out a physical model on a
cubic lattice, which is embedded in a finite linear space over a Galois field.

Recently the use of discrete space-time variables in Lattice Gauge Theories have become very
popular to construct models based on experimental date. An other succesful line of research
deals with integrable models on discrete space-time.

Many other hypothesis can be mentioned but they are reduced to two types: i) those using the
space-time lattice as an artificial model, and ii) those proposing the discreteness of
space-time as some fundamental structure of reality (Lorente 1986c).

The first assumption is very old and of technical character and do not present conceptual
problems. The second assumption is more recent, although is based on some old
pressupositions, namely, the relational character of the structure of space and time in
contradistinction to the absolute idea of space and time (Earman 1989). Both assumptions
were defended in XVII century by Leibniz and Newton, respectively, and although the last one
become more succesful the former remained as a consistent model (Jammer 1969). Acording to
Leibniz ``Time is the order of monads not existing simultaneously. Space is the order or
monads that coexist or exist simultaneously. A point changes its position when it changes
its relations from some points to different ones'' \footnote {$^\dagger$}{\tenrm According to Jammer,
Leibniz's Monadology was inspired by the philosopher Maimonides who in his {\it Guide for the
Perplexed} describe some discrete structure of space and time.}. This assumption leads to some
discrete structure, although Leibniz did not carried out all its consequences. Also Riemann is his
1854 Inaugural Dissertation discussed the posibility of a discrete manifold based on some intrinsic
metric and Weyl took over Riemann's ideas (Gr\"unbaum 1977). The relational character of space and
time have been advocated recently by R. Penrose (1971), D. Finkelstein and C.F. von Weisz\"acker
(Castell 1986). We have also proposed some physical model based on the relational hypothesis of
the space and time (Lorente 1974, 1976, 1986a,b, 1987).

In this paper we explore some mathematical consequences of the assumption of a discrete
space-time in his most naive structure, namely, a hypercubic lattice. Obviously the
mathematical tool must be the functions of discrete variables and difference equations.
Therefore, the fundamental laws of physics have to be translated into the language of discrete
mathematics. The objection which is usually raised against such discrete schemes is that they
are not invariant under the Lorentz group. We try to overcome this difficulty, as Schild did
(Schild 1949), finding all integral transformations that leave the lattice of any dimension
invariant, and applying these transformations to the fundamental field equations.

In section 2 and 3, we introduce the Cayley parametrization of the classical groups with some
easy examples that can be enlarged to all semisimple Lie groups. In section 4 we review the
isomorphims between real forms and its explicit calculation. In section 5 we describe the
method to calculate all integral transformations that leave a hypercubic lattice invariant,
using the results of sections 2 to 4. In section 6 and 7 we construct a lagrangian  formalism for
the Klein Gordon and Dirac field on the lattice. In section 8 we introduce the generators of
the space and time displacement and Lorentz boost on the lattice to check the Lorentz
invariance of the model and we ``integrate'' these generators via some ``Taylor expansion".
From the physical point of view these schemes have a twofold interpretation. Either we take
the continuous limit in order to recover the continuous character of physical laws or we keep
the formulas as they appear as a consequence of the hypothesis of a discrete space-time.

\vskip 1.5truecm  {\noindent \bf 2. CAYLEY'S RATIONAL PARAMETRIZATION OF SEMISIMPLE \break GROUPS}

\medskip Let $\cal S$ be a semisimple Lie group of complex matrices $S$, which leaves
invariant some non-degenerate bilinear form. We call a matrix $S$ of the semisimple group $\cal S$
non-exceptional if det $(E+S) \not = 0$, where $E$ is the unit matrix. Cayley (1846) has proved that
every non-exceptional matrix $S$ can be expressed as follows  $$S=(E+X)^{-1}(E-X)=(E-X)(E+X)^{-1}\eqno
(2.1)$$  where $X$ is also a non-exceptional matrix.

If $G$ is the coefficient matrix of the non-degenerate bilinear form, which is left
invariant under the group $\cal S$, the non-exceptional matrices $S$ satisfy the relation
$$S^\dagger G S = G \eqno (2.2)$$ 
and because of (2.1) the corresponding matrices $X$ will also satisfy 
$$X^\dagger G + GX = 0 \eqno (2.3)$$ 

In order to obtain the independent parameters of the semisimple group $\cal S$ it is more
convenient to work with expression (2.3), which is linear, rather than with expression
(2.2), which is quadratic. If we diagonalize or reduce to the canonical form the
coefficient matrix $G$ we have a further simplification of (2.3). Taking the independent
elements of the matrix $X$ given by (2.3) to be the independent parameters, we obtain
Cayley's rational parametrization of the semisimple group $\cal S$. (Note that when the
independent elements of $X$ are complex their real and imaginary part should be taken as
independent parameters.)

In Table 1 we give the explicit conditions on the non-exceptional matrices $S$ and $X$ for
all classical groups, as derived from expression (2.2) and (2.3), respectively. The
notation $S^T$ means the transpose matrix and $S^\dagger$ the adjoint. Also $$\vert{E+S}\vert \equiv
{\rm det} (E+S)$$
and
 $$J \equiv \pmatrix{ 
0  &E_n  \cr 
E_n  &0 \cr 
} \qquad {\rm and} \qquad I \equiv \pmatrix{ 
E_p  &0  \cr 
0  &-E_q \cr }$$ 
where $E_n, E_p, E_q$ are the unit matrix of order $n, p, q$ respectively. The condition on
the matrix $X$ gives automatically the unimodularity condition 
$${\rm det} (E+S) = {\rm det}(E-S)\eqno (2.4)$$ 
except in the groups $SU(n+1)$ and $SU(p,q)$ and therefore (2.4) imposes an extra condition
on the parameters corresponding to these groups.

$$\vbox{ \offinterlineskip  
{\centerline {TABLE 1. Cayley's decomposition of semisimple groups}}
\vskip 2 mm
\hrule
\openup 2 mm
\halign {#\  &#\  &#\  &#\quad &# \cr
\noalign{\smallskip}
&Conditions  &Conditions  &  &  \cr
Group  &on $S$ \hfill  &on $X$ \hfill &Unimodularity  &Parameters  \cr
\noalign{\smallskip}
\noalign{\hrule}
\noalign{\smallskip}
$SO(2n)$  &$S^TS=E$\hfill  &$X^T+X=0$\hfill  &  &$n(2n-1)$\hfill  \cr
$SO(2n+1)$  &$S^TS=E$\hfill  &$X^T+X=0$\hfill  &  &$n(2n+1)$\hfill  \cr
$SU(n+1)$  &$S^\dagger S=E$\hfill  &$X^\dagger +X=0$\hfill  &$\vert E+S \vert =\vert E-S
\vert$\hfill  &$n(n+2)$ \hfill \cr 
$Sp(2n)$  &$S^TJS=J$\hfill  &$X^TJ+JX=0$\hfill  & 
&$n(2n+1)$\hfill  \cr 
$SO(p,q)$  &$S^TIS=I$\hfill  &$X^TI+IX=0$\hfill  & 
&$1/2(p+q)(p+q-1)$\hfill \cr 
$SU(p,q)$  &$S^TIS=I$\hfill  &$X^TI+IX=0$\hfill  &$\vert E+S \vert
= \vert E-S \vert$\hfill  &$(p+q)^2-1$ \hfill \cr
 } \vskip 2 mm \hrule}$$

\vskip 1.5truecm  {\noindent \bf 3. SOME EXAMPLES}
\vskip 1truecm  {\noindent \bf 3.1. The Rotation Group, $SO(3)$}

\medskip From Table 1 the matrix $X$ is antisymmetric and it can be
expressed in the following way 
$$X = {1 \over m} \pmatrix { 
0  &n &-p  \cr 
n  &0 &q  \cr 
p  &-q  &0 \cr } \eqno (3.1.1)$$ 
where $n, p, q$ are independent parameters and $m$ has been
introduced for convenience. Using (3.1.1.) one obtains the Cayley parametrization ot the
non-exceptional matrix of the rotation group 
$$ \displaylines{
S = {1 \over {m^2 + n^2 + p^2 + q^2}} \hfill \cr
\hfill \times \pmatrix {
{m^2-n^2-p^2+q^2}  &{-2mn+2pq}  &{2mp+2nq} \cr 
{2mn+2pq}  &{m^2-n^2+p^2-q^2}  &{-2mq+2np} \cr
{-2mp+2nq} &{2mq+2np}  &{m^2+n^2-p^2-q^2}  \cr } (3.1.2)}$$

If we define $\alpha=m+in,\quad \beta=p-iq$ and then impose $m^2+n^2+p^2+q^2=1$, the
parametrization of the matrix $S$ given (3.1.2) is identical with the parametrization used
by Wigner (1959) for the 3-dimensional rotation group. The parameters $\alpha$ and $\beta$
used by him are related to the parametrization of $SU(2)$, the covering group of $SO(3)$,
in this way 
$$S = \pmatrix { 
\alpha  &\beta \cr 
\beta^* &\alpha^* \cr }, \qquad {\vert\alpha\vert}^2 +{\vert\beta\vert }^2 =1 \eqno(3.1.3)$$

In terms of the components of the axis of rotation $(a_1, a_2, a_3)$ and of the angle of
rotation $\phi$, the Cayley parameters have the following geometrical interpretation 
$${q \over {a_1}} = {p \over {a_2}} = {q \over {a_3}} \quad , \qquad 
\cos \phi = {{m^2-n^2-p^2-q^2} \over {m^2+n^2+p^2+q^2}} \eqno (3.1.4) $$

Similar parametrization and corresponding properties can be obtained for the
$N$-dimensional rotation groups.

\vskip 1truecm  {\noindent \bf 3.2. The Unitary Group $SU(2)$}

\medskip From Table 1 the matrix $X$ is antihermitian and it can be
expressed as  
$$X = {1 \over l}\pmatrix { 
ia  &\rho \cr 
-\rho^*  &ib  \cr } \eqno (3.2.1)$$
where $a$ and $b$ are real parameters, $\rho = r + is$ and $l$ has been added for convenience.

From (2.5) and (3.2.1) one obtains 
$$S = {1 \over \Delta} \pmatrix {
l^2+ab-{\vert \rho \vert}^2+i2lb &-2l\rho \cr 
2l\rho^*  &l^2+ab-{\vert \rho \vert}^2+i2la  \cr } \eqno (3.2.2)$$ 
with $\Delta =l^2-ab+{\vert \rho \vert}^2+il^2(a+b)$.

The antihermiticity of $X$ gives 
$${\rm det}\ (E+X)^* = {\rm det}\ (E-X)$$ 
but it does not imply the unimodularity condition. (In the rotation group the antisymmetry
of $X$ does imply the unimodularity of $S$.). If we impose det $S$ = 1, from (2.1) follows
that 
$${\rm det}\ (E+X) = {\rm det}\ (E-X)$$

Both conditions, unitarity and unimodularity of $S$, give ${\rm det}(E+X)$=real, or
$$a+b={\rm Tr}\ X = 0 \eqno (3.2.3)$$ 

Substituting (3.2.3) in (3.2.2) we obtain the general expression for the unitary unimodular
matrices in two dimensions 
$$S = {1 \over \Delta} \pmatrix { 
l^2-a^2-r^2-s^2-i2la  &-2lr-i2ls \cr 
2lr-i2ls  &l^2-a^2-r^2-s^2+i2la \cr } \eqno (3.2.4)$$
with $\Delta = l^2+a^2+r^2+s^2$. Obviously the matrix (3.2.4) is equivalent to (3.1.3), but
uses different parametrization.

\vskip 1truecm  {\noindent \bf  3.3. The Proper Lorentz Group SO(3.1) }

\medskip From Table 1 one obtains the traceless matrix 
$$X = {1 \over m} \left \{ \matrix{
0  &n  &-p  &r  \cr
-n  &0  &q  &s  \cr
p  &-q  &0  &t  \cr
r  &s  &t  &0  \cr
} \right \} \eqno (3.3.1)$$
where $n, p, q, r, s, t$ are real independent parameters and $m$ has been introduced as
before. The unimodularity of $S$ does not impose further conditions on these parameters.
From (2.1) one gets 

$$ \displaylines{
S={1 \over \Delta }\left[{\matrix{{m}^{2}-{n}^{2}-{p}^{2}+{q}^{2}+{r}^{2}-{s}^{2}-{t}^{2}+{\lambda }^{2}\cr
2mn+2pq+2rs-2\lambda t\cr
-2mp+2nq+2rt+2\lambda s\cr
-2mr-2ns+2pt-2\lambda q\cr}}\right. \hfill \cr}$$
$$\matrix{
 -2mn+2pq+2rs+2\lambda t  & 2mn+2nq+2rt-2\lambda s\cr
{m}^{2}-{n}^{2}+{p}^{2}-{q}^{2}-{r}^{2}+{s}^{2}-{t}^{2}+{\lambda }^{2}  &-2mq+2np+2st+2\lambda r\cr
2mq+2np+2st-2\lambda r  &{m}^{2}+{n}^{2}-{p}^{2}-{r}^{2}-{s}^{2}+{t}^{2}+{\lambda }^{2}\cr
-2ms+2nr-2qt-2\lambda p  &-2mt-2pr+2qs-2\lambda n  \cr} $$ 
$$\eqalignno{ \hfill \left.{ \matrix{
 -2mr+2ns-2pt-2\lambda q  \cr
-2ms-2nr+2qt-2\lambda p  \cr
-2mt+2pr-2qs-2\lambda n  \cr
{m}^{2}+{n}^{2}+{p}^{2}+{q}^{2}+{r}^{2}+{s}^{2}+{t}^{2}+{\lambda }^{2}  \cr}
}\right] (3.3.2) \cr}$$
where 
$$
m\lambda =nt+ps+qr$$
and
$$\Delta =m^2+n^2+p^2+q^2-r^2-s^2-t^2-\lambda^2 $$

If $\Delta > 0$, since det $S = 1$, one obtains the general expression for
the non-exceptional matrices of the proper Lorentz group $(S_{44} > 0)$.

If $r = s = t = 0$, one recovers expression (3.1.2) for the proper orthogonal group in
3-dimensions.

If $n = p = q = 0$ one is left with the non-exceptional matrices of the pure Lorentz
transformations. In this case, comparison of (3.3.2) with a pure Lorentz transformation
with velocity $v$ along {\bf v} gives (M\o ller, 1952) 
$${v_x \over r}={v_y \over s} ={ v_z \over t }= {2mc \over {m^2+r^2+s^2+t^2}}, \qquad \left
(1-{v^2 \over c^2}\right )^{1/2}={{m^2-r^2-s^2-t^2}\over{m^2+r^2+s^2+t^2}} \eqno (3.3.3)$$ 
where $c$ is the velocity of light in vacuum.

If we define
$$\left.{\matrix{ \alpha =m-t+i\left({n-\lambda }\right),&\beta =-p-r+i\left({q-s}\right)\cr
\gamma =p-r+i\left({q+s}\right),&\delta =m+t-i\left({n+\lambda }\right)\cr}}\right\} \eqno
(3.3.4)$$
and introduce these variables in the general expression of the proper Lorentz group
in terms of the parameters of the $SL (2, C)$ group (Naimark, 1964a)
$$\left({\matrix{\rm \alpha &b\cr
\gamma &\delta \cr}}\right) \qquad \alpha \delta -\beta \gamma =1$$
we obtain the expression (3.3.2) plus the condition
$$\left.{\matrix{
m\lambda &=nt+ps+qr  \hfill \cr
\Delta &={m}^{2}+{n}^{2}+{p}^{2}+{q}^{2}-{r}^{2}-{s}^{2}-{t}^{2}-{\lambda}^{2}=1 \hfill \cr}}\right\}
\eqno(3.3.5)$$

\vskip 1.5truecm  {\noindent \bf 4. ISOMORPHISM BETWEEN REAL FORMS }

\medskip According to Cartan theory, there are some real forms of simple Lie groups of low
dimensionality which are locally isomorphic (Helgason, 1978). We describe them by the
bijection of $R^n$ onto a set of matrices $S$.

\medskip i) $SL(2,R) \approx SO(2,1)$. Define a set of 2$\times$2 real matrices $A$, by the
conditions $A^T = A$, where $A^T$ means transposed. The bijection of an element
$(x_0,x_1,x_2)$ of $R^3$ onto a matrix $A$ is the following:
$$ A=\left({\matrix{{x}_{0}+{x}_{2}&{x}_{1}\cr
{x}_{1}&{x}_{0}-{x}_{2}\cr}}\right) \eqno (4.1)$$

The transformations $A' = SAS^T$ with $S\in SL(2,R)$, map $A$ into itself. Since
$${\rm det}\ A={x}_{0}^{2}-{x}_{1}^{2}-{x}_{2}^{2}={\rm det}\  A' \eqno (4.2)$$
this transformation induces the desired isomorphism.

\medskip (ii) $SL(2,C)\approx SO(3,1)$. Define $A$, a 2$\times$2 complex matrix, by the
condition $A^\dagger = A$, where $A^\dagger$ means the Hermitian conjugate matrix. The
bijection of $(x_0,x_1,x_2,x_3)$ in $R^4$ onto $A$ is given by
$$A=\left({\matrix{{x}_{0}+{x}_{3}&{x}_{1}+i{x}_{2}\cr
{x}_{1}-i{x}_{2}&{x}_{0}-{x}_{3}\cr}}\right) \eqno (4.3)$$

The transformation $A' = SAS^\dagger$ with $S\in SL(2,C)$ maps $A$ into itself, as it
is well known (Gel'fand et al., 1963). Since
$${\rm det}A={x}_{0}^{2}-{x}_{1}^{2}-{x}_{2}^{2}-{x}_{3}^{2}={\rm det}A' \eqno (4.4)$$
this transformation induces the mentioned isomorphism.

\medskip (iii) $Sp(4,R)\approx SO(3,2)$. The matrix $A$ is a four-dimensional real matrix, satisfying
$A^TJ = JA$ and $Tr A = 0$, where
$$ J\equiv \left({\matrix{0  &E\cr
-E  &0\cr}}\right),$$
$E$ is the unit matrix of dimension 2. 

The bijection of an element $(x_1,x_2,x_3,x_4,x_5)$, of $R^5$ onto $A$ is given by
$$A=\left({\matrix{
{x}_{1}  &{x}_{2}+{x}_{3}  &0  &{x}_{4}+{x}_{5}  \cr
{x}_{2}-{x}_{3}  &-{x}_{1}  &{-x}_{4}-{x}_{5}  &0  \cr
0  &{x}_{4}-{x}_{5}  &{x}_{1}  &{x}_{2}-{x}_{3}  \cr
{-x}_{4}+{x}_{5}  &  &{x}_{2}+{x}_{3}  &-{x}_{1}  \cr
}}\right) \eqno (4.5)$$

The transformation $A' = SAS^{-1}$ with $S\in Sp(4,R)$ maps $A$ into itself, namely, $A'^TJ =
JA', TrA' = 0$. Since
$${\rm
det}\ A={\left({{x}_{1}^{2}+{x}_{2}^{2}-{x}_{3}^{2}
-{x}_{4}^{2}+{x}_{5}^{2}}\right)}^{2}={\rm det}A',\eqno(4.6)$$ 
this transformation induces the desired isomorphism.

\medskip (iv) $Sp(1,1)\approx SO(4,1)$ $A$ is defined by the four-dimensional complex matrix satisfying
$A^TJ = JA,\quad A^\dagger K = KA, \quad Tr\ A = 0$, with $K\equiv \rm{diag}(1,-1,1,-1)$. 

The bijection of an element $(x_1,x_2,x_3,x_4,x_5)$ of $R^5$ onto $A$ is
$$A=\left({\matrix{{x}_{1}&{x}_{2}+{ix}_{3}&0&{x}_{4}+{ix}_{5}\cr
{-x}_{2}+{ix}_{3}&-{x}_{1}&{-x}_{4}-{ix}_{5}&0\cr
0&{x}_{4}-{ix}_{5}&{x}_{1}&{-x}_{2}+{ix}_{3}\cr
{-x}_{4}+{ix}_{5}&0&{x}_{2}+{ix}_{3}&{-x}_{1}\cr}}\right) \eqno (4.7)$$

Given an element $S$ of the group $Sp(1, 1)$, that is to say, $S^TJS = J, S^\dagger KS = K$,
the transformation $A' = SAS^{-1}$ maps $A$ into itself. Since
$${\rm det}\
A={\left({{x}_{1}^{2}-{x}_{2}^{2}-{x}_{3}^{2}-{x}_{4}^{2}-{x}_{5}^{2}}\right)}^{2}= {\rm
det}\ A' \eqno (4.8)$$ 
this transformation induces the desired isomorphism.

\medskip (v) $SU(2,2)\approx SO(4,2)$. $A$ is defined by the four-dimensional complex matrix,
satisfying $A^T=-A,\quad A^*I=I\bar A $, with $\bar A$, the complex conjugate matrix of $A,
A^*$ the dual matrix of $A$, namely, $\left({\rm A}^* \right)_{ab}={1 \over 2}{\varepsilon
}_{abcd}{A}^{cd}$, and 
$$\rm I\equiv \left({\matrix{
E  &0\cr
0  &-E\cr}}\right)$$

The bijection of an element $(x_1,x_2,x_3,x_4,x_5,x_6)$ of $R^6$ onto $A$ is (Beckers et al.,
1978) $$ A=\left({\matrix{0&{x}_{1}+{ix}_{2}&{x}_{3}+{ix}_{4}&{x}_{5}+{ix}_{6}\cr
{-x}_{1}-{ix}_{2}&0&{x}_{5}-{ix}_{6}&{-x}_{3}+{ix}_{4}\cr
{-x}_{3}-{ix}_{4}&{-x}_{5}+{ix}_{6}&0&{-x}_{1}+{ix}_{2}\cr
{-x}_{4}-{ix}_{6}&{x}_{3}-{ix}_{4}&{x}_{1}-{ix}_{2}&0\cr}}\right) \eqno (4.9)$$

The transformation $A' = SAS^T$, with $S$ satisfying $S^\dagger IS = I$, maps $A$ into itself.
Since
$${\rm
det}\
A={\left({{-x}_{1}^{2}-{x}_{2}^{2}+{x}_{3}^{2}+{x}_{4}^{2}+{x}_{5}^{2}+{x}_{6}^{2}}\right)}={\rm
det}A' \eqno (4.10)$$ this transformation belongs also to $SO(4, 2)$.

\medskip (vi) $SL(2,Q)\approx SO(5,1)$. Let $A$ be a two-dimensional quaternion matrix defined by
$A^\dagger = A$. The bijection of an element $(x_0,x_1,x_2,x_3,x_4,x_5)$ of $R^6$ onto $A$ is
the following: 
$$ A=\left({\matrix{{x}_{0}+{x}_{1}&{x}_{2}+{x}_{3}i+{x}_{4}j+{x}_{5}k\cr
{x}_{2}-{x}_{3}i-{x}_{4}j-{x}_{5}k&{x}_{0}-{x}_{1}\cr}}\right) \eqno (4.11)$$
with $(i,j,k)$ a basis for the quaternions. The transformation $A' = SAS^\dagger$, with
$S\in SL(2,Q)$ maps $A$ into itself. Since.
$${\rm
det}\ A={\left({{x}_{0}-{x}_{1}^{2}-{x}_{2}^{2}-{x}_{3}^{2}
-{x}_{4}^{2}-{x}_{5}^{2}}\right)}^{2}={\rm det}A' \eqno (4.12)$$ 
this transformation induces the desired isomorfhism (Barut et al., 1965)

\medskip (vii) $SL(4,R)\approx SO(3,3)$. $A$ is defined by the four-dimensional real matrix,
satisfying $A^T = -A$. The bijection of an element $(x_1,x_2,x_3,x_4,x_5,x_6)$ of $R^6$ onto
$A$ is the following: 
$$
A=\left({\matrix{0&{-x}_{1}+{x}_{4}&{x}_{2}+{x}_{5}&{x}_{3}+{x}_{6}\cr
{-x}_{1}-{x}_{4}&0&{x}_{3}-{x}_{6}&{-x}_{2}+{x}_{5}\cr
{-x}_{2}-{x}_{5}&{-x}_{3}+{x}_{6}&0&{x}_{1}-{x}_{4}\cr
{-x}_{3}-{x}_{6}&{x}_{2}-{x}_{5}&{-x}_{1}+{x}_{4}&0\cr}}\right) \eqno (4.13)$$

The transformation $S' = SAS^T$, with $S\in SL(4,R)$, maps $A$ into itself. Since
$${\rm
det}\ A={\left({{x}_{1}^{2}+{x}_{2}^{2}+{x}_{3}^{2}-{x}_{4}^{2}-{x}_{5}^{2}
-{x}_{6}^{2}}\right)}^{2}={\rm det}A' \eqno (4.14)$$ 
this transformation induces the desired isomorphism.

\ 

\ 

\vskip 1.5truecm {\noindent \bf 5. INTEGRAL TRANSFORMATIONS} 

\medskip A transformation that leaves invariant an hypercubic lattice is called an integral
transformation. It means that if we take an integral vector (the components of which are
real integers or Gaussian numbers) the transformed vector is also integral. 

It is easy to prove that a transformation of a classical group is integral if and only if
all its components are integers.

Given the Cayley realization of a classical groups (2.1) the corresponding transformation
is integral if all the Cayley parameters are integers and det $(1 + X) = 1$. But this
procedure do not exhaust all the integral transformations. Let us give some examples.

\vskip 1truecm {\noindent \bf 5.1 $SO(3)$} 

\medskip The Cayley parametrization is given by (3.1.2.). Therefore we require $m, n, p, q$
to be integer numbers and 
$$m^2 + n^2 + p^2 + q^2 = 1  \eqno (5.1.1)$$ 

The non-exceptional matrix satisfying both conditions is the unit matrix corresponding to
$m = \pm 1 ,\quad n = p=q = 0$

The other set of non trivial integral transformation are obtained in this way: from the
Cayley transform
$$S={1-X \over 1+X}={2 \over 1+X}-1 \eqno (5.1.2)$$
it follows that $S$ is integral if $2(1 + X)^{-1}$ is integral.

From (3.1.2) we get:
$${2 \over 1+X}={2
\over{m}^{2}+{n}^{2}+{p}^{2}+{q}^{2}}\left({\matrix{
{m}^{2}+{q}^{2}  &-mn+pq  &mp+nq  \cr
mn+pq  &{m}^{2}+{p}^{2}  &-mq+np  \cr 
-mp+nq  &mq+np  &{m}^{2}+{n}^{2}  \cr}}\right) \eqno(5.1.3)$$

Obviously $2(1+X)^{-1}$ is integral if $m, n, p, q$ are integers and $m^2 + n^2 + p^2 +
q^2 = 2$. The solutions of this diophantine equation are:
$$\left.{\matrix{
m=\pm 1 &, \quad n=\pm 1  &, \quad p=q=0  \cr
m=\pm 1 &, \quad p=\pm 1  &, \quad n=q=0  \cr
m=\pm 1 &, \quad q=\pm 1  &, \quad n=p=0  \cr}}\right\} \eqno(5.1.4)$$
By similar considerations we obtain also the integral transformations with
$$m=\pm 1 \quad, \quad n=\pm 1  \quad, \quad p=\pm 1  \quad, \quad q=\pm 1 \eqno(5.1.5)$$.

 {\noindent \bf 5.2 $SO(3,1)$} 

\medskip  As before we require in (3.3.2) all the Cayley parameters $m, n, p, q, r, s,
t$ to be integer numbers; after substituting $\lambda = (nt + ps + qr)/m$ the condition
det $(1 + X) = 1$ becomes
$$m^2(m^2+n^2+p^2+q^2-r^2-s^2-t^2)-(nt+ps+qn)^2=m^2 \eqno (5.2.1)$$

This formidable diophantine equation can be simplified if we expand the inverse matrix
appearing in the Caley transform (5.1.2), namely,
$${1 \over 1+X}=1-X+X^{2}-X^{3}+\cdots$$

This series must be finite if matrix $S$ is supposed to be an integral transformation,
therefore ${\rm X}^r = 0$ for some $r$. Since $X$ is a nihilpotent matrix, it has zero as the
only eigenvalue. From the secular equation it follows that the sum of all principal minors of the
same order of the matrix $X$ are equal to zero. This property applied to (3.3.1) gives the conditions
$$ \left \{ \matrix {
n^2+p^2+q^2-r^2-s^2-t^2 = 0 \cr
(nt+ps+qr)^2  = 0 \cr} \right . \eqno (5.2.2)$$
substituting these values in (5.2.1) we get $m =1$.

An other set of solutions are obtained from the condition $2(1 + X)^{-1}$ to be integral.

This condition together with (5.2.2) gives
$$\displaylines{
{ 2 \over 1+X}={2 \over {m}^{2}} \times \hfill \cr \cr
\hfill  \pmatrix{
{m}^{2}+{q}^{2}-{s}^{2}-{t}^{2}  &-mn+pq+rs  &mp+nq+rt  &-mr=ns-pt  \cr
mn+pq+rs  &{m}^{2}+{p}^{2}-{r}^{2}-{t}^{2}  &-mq+np+st  &-ms-nr+qt  \cr
-mp+nq+rt  &mq+np+st  &{m}^{2}+{n}^{2}-{r}^{2}-{s}^{2}  &-mt+pr-qs  \cr
-mr-ns+pt  &-ms+nr-qt  &-mt-pr+qs  &{m}^{2}+{n}^{2}+{p}^{2}+{q}^{2}  \cr} \cr
\hfill(5.2.3)\cr} $$

The condition of integral transformation requires that each matrix element multiplied by 2 must
be divisible by $m^2$. Therefore the only solutions are: 
$$ \left . \matrix { 
&m=2 \quad , &n,t,p,s  &{\rm odd \quad integers;} &q,r &{\rm even \quad integers} \cr 
{\rm or} &&&&&\cr 
&m=2 \quad , &n,t,q,r  &{\rm odd \quad integers;} &p,s &{\rm even \quad integers} \cr} 
\right \} \eqno (5.2.4)$$

The same method can be applied to other integral transformations corresponding to classical
groups.

\vskip 1truecm {\noindent \bf 5.3. The method of isomorphism between real forms} 

\medskip  Section 4 can be used to find integral transformations. Take the isomorphism
between $SL(2,C)$ and $SO(3,1)$. A spin transformation of the group $SL(2,C)$ corresponds
to a Lorentz tansformation with Cayley parametrization after the identification (3.3.4).

An spin transformation is integral if the corresponding Lorentz transformation is integral. 
Schild has solved completely the problem of classifying all integral Lorentz tansformation
by the following theorem (Schild 1949):

A spin transformation $\lambda_{ij} \quad (i,j=1,2)$ is integral if and only if one of the following
four conditions is satisfied:

\smallskip \begingroup \parindent=1cm
\item{I.} $\lambda_{ij}$ are Gaussian integers such that $\lambda_{11}\lambda_{22} -
\lambda_{21}\lambda_{12} = 1$ and such that $\lambda_{11} + \lambda_{21} + \lambda_{12} +
\lambda_{22}$ is even.

\item{II.} $\lambda_{ij} = \mu_{ij}/(1 + i)$, 
where $\mu_{ij}$ are odd integers such that $\mu_{11}\mu_{22}-\mu_{21}\mu_{12} = 2i$ 

\item{III.} $\lambda_{ij}$ are integers such that $\lambda_{11}\lambda_{22} -
\lambda_{21}\lambda_{12} = i$ and such that $\lambda_{11} + \lambda_{12} + \lambda_{21} +
\lambda_{22}$ is even.

\item{IV.} $\lambda_{ij} = \mu_{ij}/(1 + i)$
where $\mu_{ij}$ are odd integers such that $\mu_{11}\mu_{22} - \mu_{12}\mu_{21} = -2$
 
(For the definition and properties of Gaussian numbers see Schild 1949).
\par \endgroup \smallskip

We can compare this classification with our results of section 5.2. It is easy to check
with the help of (3.3.4) that Case I corresponds to $m = 1$ and (5.2.2) and Case II
corresponds to $m = 2$ and (5.2.4).

Cases III an IV are obtained from I and II multiplying $\alpha, \beta$ or $\gamma, \delta$
by $i$ in (3.3.4). It corresponds to interchange first and second row in (3.3.2), multiplying the
first one by -1. If we choose to multiply $\alpha, \gamma$ or $\beta, \delta$ by $i$ in (3.3.4) we obtain
the corresponding transformation (3.3.2) with the first and second column (multiplied by $-1$) 
interchanged.

In order to complete the algorithm to calculate the $SL(2,C)$ transformations with Gaussian numbers,
we have to solve the diophantine equation $\lambda_{11}\lambda_{22} - \lambda_{12}\lambda_{21}
= 1$. As before we write the general element of $SL(2,C)$ as a finite series, say
$$S={1 \over 1+X}=1-X \qquad {\rm with } \quad X^2=0  \eqno (5.3.1)$$
the condition for $X$ to be a nihilpotent matrix requires
$${\rm det} {X}=0 \qquad {trX}=0 \quad , \eqno (5.3.2)$$
the most general solution of which is
$$ { X}=\left({\matrix{
zw,  &-z{w}^{2}  \cr
z,  &-zw  \cr}}\right) \eqno (5.3.3)$$
with $z, w$ Gaussian integers. Therefore
$$S=\left({\matrix{
1-zw ,  &-zw^{2}  \cr
z,  &1+zw   \cr}}\right) \equiv \left({\matrix{
{\lambda }_{11}  &{\lambda }_{12}  \cr
{\lambda }_{21}  &{\lambda }_{22}  \cr}}\right) \eqno (5.3.4)$$
that can be enlarged to the general solution
$$S=\left({\matrix{
{\lambda }_{11}+p{\lambda }_{21}  &{\lambda }_{12}+p{\lambda }_{22}  \cr
{\lambda }_{21}  &{\lambda }_{22}  \cr}}\right) \eqno (5.3.5)$$
with $p$, an arbitrary Gaussian integer.

\vskip 1.5truecm {\noindent \bf 6. A LAGRANGIAN MODEL FOR THE KLEIN-GORDON FIELD ON THE
LATTICE} 

\medskip  Let us define the real scalar field $\phi (x, t)$ on the grid points of a (1 +
1)-dimensional lattice as $\phi (\varepsilon j, \tau n) \equiv \phi ^n_j$ where $\varepsilon,
\tau$  are the fundamental space and time interval and $j,n$ are integer numbers. We want to
associate to this field a Lagrangian, such that the equations of motion are recovered from
the Euler-Lagrange difference equations. 

A suitable Lagrangian for the Klein-Gordon field is: 
$$\displaylines{
{L}_{n}=-\sum\nolimits\limits_{j=0}^{N-1} {1 \over 2}\left\{{{1 \over {\varepsilon
}^{2}}{\left({{\tilde{\nabla }}_{j}{\Delta }_{j}{\tilde{\nabla }}_{n}{\phi
}_{j}^{n}}\right)}^{2}-{1 \over {\tau }^{2}}{\left({{\tilde{\nabla }}_{j}{\tilde{\Delta
}}_{j}{\nabla }_{n}{\phi }_{j}^{n}}\right)}^{2}+{M}^{2}{\left({{\tilde{\nabla
}}_{j}{\tilde{\Delta }}_{j}{\tilde{\nabla }}_{n}{\phi }_{j}^{n}}\right)}^{2}}\right\} \cr
\hfill \equiv \varepsilon\sum\nolimits\limits_{j=0}^{N-1} {\cal L}_{n} \qquad (6.1) \cr}$$
where periodic boundary conditions for the field are supposed, $\Delta_j (\nabla_j)$ are the
forward (backward) difference operator and $\tilde \Delta_j(\tilde \nabla_j)$ are the
forward (backward) average operator with respect to the space indices:
$${ \Delta }_{j}{f}_{j}\equiv {f}_{j+1}-{f}_{j}\qquad {\nabla
}_{j}{f}_{j}={f}_{j}-{f}_{j-1} \eqno (6.2)$$ 
$${\tilde{\Delta }}_{j}{f}_{j}={1 \over
2}\left({{f}_{j+1}+{f}_{j}}\right) \qquad {\tilde{\nabla }}_{j}{f}_{j}={1 \over
2}\left({{f}_{j}+{f}_{j-1}}\right) \eqno (6.3)$$
and similar for the time indices. In the limit $j \rightarrow \infty ,
\quad\varepsilon\rightarrow 0 , \quad j\varepsilon \rightarrow
x,\quad n\rightarrow \infty , \quad\tau \rightarrow 0,\quad
n\tau\rightarrow t$ \quad we have
$${\tilde{ \nabla }}_{j}{\tilde{\Delta }}_{j}{\tilde{\nabla }}_{n}{\phi
}_{j}^{n}\rightarrow \phi \left({x,t}\right) \eqno (6.4)$$ 
$${\rm 1 \over \varepsilon
}{\tilde{\nabla }}_{j}{\Delta }_{j}{\tilde{\nabla }}_{n}{\phi }_{j}^{n}\rightarrow {\partial \phi 
\over \partial x}\left({x,t}\right) \eqno (6.5)$$ 
$${1 \over \tau
}{\tilde{\nabla }}_{j}{\tilde{\Delta }}_{j}{\nabla }_{n}{\phi }_{j}^{n}\rightarrow {\partial
\phi  \over \partial t}\left({x,t}\right) \eqno (6.6)$$

Taking the variations of the Lagrangian density ${\cal L}_n$ with respect to the time
difference of the field, we get
$${\partial {\cal L}_{n} \over \partial \left({{1 \over \tau }{\nabla }_{n}{\tilde{\Delta
}}_{j}{\tilde{\nabla }}_{j}{\phi }_{j}^{n}}\right)}={1 \over \tau }{\nabla
}_{n}{\tilde{\Delta }}_{j}{\tilde{\nabla }}_{j}{\phi }_{j}^{n}\equiv {\tilde{\nabla
}}_{n}{\tilde{\Delta }}_{j}{\tilde{\nabla }}_{j}{\pi }_{j}^{n} \eqno (6.7)$$
with $\pi_j^n$ as the conjugate momentum.

To obtain the Euler-Lagrangian equation we take the time difference of the last expresion
to be equal to the variation of the Lagrangian density with respect to the average field:
$${\rm 1 \over {\tau }^{2}}{\Delta }_{n}{\nabla }_{n}{\tilde{\Delta }}_{j}{\tilde{\nabla
}}_{j}{\phi }_{j}^{n}={\tilde{\Delta }}_{n}{\partial {\cal L}_{n} \over \partial
\left({{\tilde{\nabla }}_{j}{\tilde{\Delta }}_{j}{\tilde{\nabla }}_{n}{\phi
}_{j}^{n}}\right)}={\tilde{\Delta }}_{n}\left[{{1 \over {\varepsilon }^{2}}{\nabla
}_{j}{\Delta }_{j}{\tilde{\nabla }}_{n}{\phi }_{j}^{n}-{M}^{2}{\tilde{\nabla
}}_{j}{\tilde{\Delta }}_{j}{\tilde{\nabla }}_{n}{\phi }_{j}^{n}}\right] \eqno (6.8)$$
where $\tilde \Delta_n$ has been introduce for homogeneity in the last equality and integration by
parts have been used. 

The last expression is the wave equation for the Klein-Gordon field on the lattice. 

The ``plane wave'' solutions, the Fourier decomposition of the fields in terms of a complete set
of orthogonal functions on the lattice have been given elsewhere (Lorente, 1992).

With the help of the conjugate field defined in (6.7) we can construct an Hamiltonian
density on the lattice in the usual way.
$${\cal H}_{n}=\left({\tilde{\nabla }}_{j}{\tilde{\Delta }}_{j}{\tilde{\nabla }}_{n}{\pi
}_{j}^{n}\right){{1 \over \tau }{\nabla }_{n}{\tilde{\Delta }}_{j}{\tilde{\nabla }}_{j}{\phi
}_{j}^{n}}-{\cal L}_{n} \eqno (6.9)$$
from which the Hamiltonian follows:
$$\displaylines{
{H}_{n}=\varepsilon \sum\nolimits\limits_{j=0}^{N-1} {1 \over
2}\left\{{{\left({{\tilde{\nabla }}_{j}{\tilde{\Delta }}_{j}{\tilde{\nabla }}_{n}{\pi
}_{j}^{n}}\right)}^{2}+{1 \over {\varepsilon }^{2}}\left({{\tilde{\nabla }}_{j}{\Delta
}_{j}{\tilde{\nabla }}_{n}{\phi }_{j}^{n}}\right)+{M}^{2}{\left({{\tilde{\nabla
}}_{j}{\tilde{\Delta }}_{j}{\tilde{\nabla }}_{n}{\phi }_{j}^{n}}\right)}^{2}}\right\} \cr
\hfill=\varepsilon\sum\nolimits\limits_{j=0}^{N-1}{\cal H}_{n} \qquad(6.10) \cr}$$

Taking the variation of the Hamiltonial density with respect to the averaged fields and its
conjugate momentum, the Hamilton equations of motion are obtained:
$${{1 \over \tau }{\Delta }_{n}\left({{\tilde{\nabla }}_{j}{\tilde{\Delta
}}_{j}{\tilde{\nabla }}_{n}{\phi }_{j}^{n}}\right)}^{}={\partial {\cal H}_{n} \over \partial
\left({{\tilde{\nabla }}_{j}{\tilde{\Delta }}_{j}{\tilde{\nabla }}_{n}{\pi
}_{j}^{n}}\right)}={\tilde{\Delta }}_{n}\left({{\tilde{\nabla }}_{j}{\tilde{\Delta
}}_{j}{\tilde{\nabla }}_{n}{\pi }_{j}^{n}}\right) \eqno (6.11)$$
$$\eqalignno{
{{1 \over \tau }{\Delta }_{n}\left({{\tilde{\nabla }}_{j}{\tilde{\Delta
}}_{j}{\tilde{\nabla }}_{n}{\pi }_{j}^{n}}\right)}^{} &=-{\partial {\cal H}_{n} \over \partial
\left({{\tilde{\nabla }}_{j}{\tilde{\Delta }}_{j}{\tilde{\nabla }}_{n}{\phi
}_{j}^{n}}\right)} \cr
&={\tilde{\Delta }}_{n}\left\{{{1 \over {\varepsilon }^{2}}{\nabla
}_{j}{\Delta }_{j}{\tilde{\nabla }}_{n}{\phi }_{j}^{n}-{M}^{2}{\tilde{\nabla
}}_{j}{\tilde{\Delta }}_{j}{\tilde{\nabla }}_{n}{\phi }_{j}^{n}}\right\} &(6.12)\cr}$$
where integration by parts has been used. Again, the Hamiltonian equations of motion lead
to the wave equation (6.8). Notice that (6.11) and (6.12) can be simplified by the time
average operator $\tilde \nabla_n$, namely:
$${{\rm 1 \over \tau }{\Delta }_{n}\left({{\tilde{\nabla }}_{j}{\tilde{\Delta }}_{j}{\phi
}_{j}^{n}}\right)}^{}={\tilde{\Delta }}_{n}\left({{\tilde{\nabla }}_{j}{\tilde{\Delta
}}_{j}{\pi }_{j}^{n}}\right) \eqno (6.13)$$
$${{\rm 1 \over \tau }{\Delta }_{n}\left({{\tilde{\nabla }}_{j}{\tilde{\Delta }}_{j}{\pi
}_{j}^{n}}\right)}^{}={\tilde{\Delta }}_{n}\left({{1 \over {\varepsilon }^{2}}{\nabla
}_{j}{\Delta }_{j}{\phi }_{j}^{n}-{M}^{2}{\tilde{\nabla }}_{j}{\tilde{\Delta }}_{j}{\phi
}_{j}^{n}}\right) \eqno (6.14)$$
which can be obtained from the following Hamiltonian
$${H}_{n}=\varepsilon \sum\nolimits\limits_{j=0}^{N-1} {1 \over
2}\left\{{{\left({{\tilde{\nabla }}_{j}{\tilde{\Delta }}_{j}{\pi }_{j}^{n}}\right)}^{2}+{1
\over {\varepsilon }^{2}}{\left({{\tilde{\nabla }}_{j}{\Delta }_{j}{\phi
}_{j}^{n}}\right)}^{2}+{M}^{2}{\left({{\tilde{\nabla }}_{j}{\tilde{\Delta }}_{j}{\phi
}_{j}^{n}}\right)}^{2}}\right\} \eqno (6.15)$$

For the quantization of the Klein-Gordon field we introduce the equal time commutation
relations:
$$\eqalignno{
&\left[{{\tilde{\rm \nabla }}_{j}{\tilde{\Delta }}_{j}{\phi }_{j}^{n},{\tilde{\nabla
}}_{j'}{\tilde{\Delta }}_{j'}{\pi }_{j'}^{n}}\right]={1 \over \varepsilon }{\delta }_{jj'}
&(6.16) \cr
&\left[{{\tilde{\rm \nabla }}_{j}{\tilde{\Delta }}_{j}{\phi }_{j}^{n},{\tilde{\nabla
}}_{j'}{\tilde{\Delta }}_{j'}{\phi }_{j'}^{n}}\right]=0=\left[{{\tilde{\nabla
}}_{j}{\tilde{\Delta }}_{j}{\pi }_{j}^{n},{\tilde{\nabla }}_{j'}{\tilde{\Delta }}_{j'}{\pi
}_{j'}^{n}}\right] &(6.17) \cr}$$
from which the Heisenberg equations of motion, the Fourier decomposition in terms of the
creation and annihilation operators can be deduced in the usual way (Lorente 1992).

\vskip 1.5truecm {\noindent \bf 7. A LAGRANGIAN FOR THE DIRAC FIELD ON THE
LATTICE} 

\medskip  We can repeat the same steps as in the previous case for the Dirac field on the
$(1+1){\rm -dimensional}$ lattice ${\rm \psi }_{\alpha }\left({j\varepsilon ,n\tau
}\right)\equiv {\rm\psi}_{\alpha j}^{n}$. 

A suitable Lagrangian density is
$${\cal L}_{n}=-{\tilde{\Delta }}_{j}{\tilde{\Delta }}_{n}{\psi }_{j}^{\dagger
n}\left\{{{\gamma }_{4}{\gamma }_{1}{1 \over \varepsilon }{\Delta }_{j}{\tilde{\Delta
}}_{n}{\psi }_{j}^{n} - i {1\over \tau}{\gamma }_{4}{\tilde{\Delta
}}_{j}{\Delta }_{n}{\psi }_{j}^{n}} +M{\gamma }_{4}{\tilde{\Delta
}_{j}{\tilde{\Delta }}_{j}{\psi }_{j}^{n}}\right\} \eqno (7.1)$$
with
$${\rm \gamma }_{1}=\left({\matrix{
0  &-i  \cr
i  &0  \cr}}\right),\qquad {\gamma }_{4}=\left({\matrix{
1  &0  \cr
0  &-1  \cr}}\right),\qquad i{\gamma }_{1} {\gamma }_{4}={\gamma }_{5}=\left({\matrix{
0  &-1  \cr
-1  &0  \cr}}\right) \eqno (7.2)$$
leading to the Euler-Lagrange equation
$${ \partial {\cal L}_{n} \over \partial \left({{1 \over \tau }{\Delta }_{n}{\tilde{\Delta
}}_{j}{\psi }_{j}^{n}}\right)}=i{\tilde{\Delta }}_{j}{\tilde{\Delta }}_{n}{\psi
}_{j}^{\dagger n}\equiv {\tilde{\Delta }}_{j}{\tilde{\Delta }}_{n}{\pi }_{j}^{n} \eqno (7.3)$$
$${1 \over \tau }{\nabla }_{n}\left({{\tilde{\Delta }}_{j}{\tilde{\Delta }}_{n}{\pi
}_{j}^{n}}\right)={\tilde{\nabla }}_{n}{\partial {\cal L}_{n} \over \partial
\left({{\tilde{\Delta }}_{n}{\tilde{\Delta }}_{j}{\psi }_{j}^{n}}\right)}={\tilde{\nabla
}}_{n}\left\{{{1 \over \varepsilon }{\Delta }_{j}{\tilde{\Delta }}_{n}{\psi }_{j}^{\dagger
n}{\gamma }_{4}{\gamma }_{1}-M{\tilde{\Delta }}_{j}{\tilde{\Delta }}_{n}{\psi }_{j}^{\dagger
n}{\gamma }_{4}}\right\} \eqno (7.4)$$

Substituting (7.3) in (7.4), taking the adjoint operation of both sides and multiplying by
$\gamma_4$ from the right we obtain the Dirac equation on the lattice
$${\tilde{\nabla }}_{n}\left\{{{\gamma }_{1}{1 \over \varepsilon }{\Delta
}_{j}{\tilde{\Delta }}_{n}-i{\gamma }_{4}{1 \over \tau }{\Delta }_{n}{\tilde{\Delta
}}_{j}+M{\tilde{\Delta }}_{j}{\tilde{\Delta }}_{n}}\right\}{\psi }_{j}^{n}=0 \eqno (7.5)$$

The plane wave solutions and the Fourier decomposition in terms of a complete set of orthogonal
functions on the lattice have been given elsewhere (Lorente 1991).

With the help of the conjugate field $\pi^n_j = i \psi_j^{\dagger n}$ we can construct the
Hamiltonian density
$$\eqalignno{
{\cal H}_{n} &=\left({{\tilde{\Delta }}_{n}{\tilde{\Delta }}_{j}{\pi }_{j}^{n}}\right){1 \over
\tau }{\Delta }_{n}{\tilde{\Delta }}_{j}{\psi }_{j}^{n}-{\cal L}_{n}  \cr
&=\left({{\tilde{\Delta
}}_{n}{\tilde{\Delta }}_{j}{\psi }_{j}^{\dagger n}}\right){\gamma }_{4}{\gamma
}_{1}{\tilde{\Delta }}_{n}{\Delta }_{j}{\psi }_{j}^{n}+M\left({{\tilde{\Delta
}}_{n}{\tilde{\Delta }}_{j}{\psi }_{j}^{\dagger n}}\right){\tilde{\Delta }}_{n}{\tilde{\Delta
}}_{j}{\psi }_{j}^{n} &(7.6) \cr}$$
from which the Hamiltonian equations of motion can be derived leading again to the Dirac
equation.

As in the Klein-Gordon case, one can simplify the Hamilton equation by $\tilde \Delta_n$ which can
be deduced from the new Hamiltonian
$${ H}_{n}=\varepsilon \sum\nolimits\limits_{j=0}^{N-1} {\tilde{\Delta }}_{j}{\psi
}_{j}^{\dagger n}\left\{{{1 \over \varepsilon }{\gamma }_{4}{\gamma }_{1}{\Delta }_{j}{\psi
}_{j}^{n}+M{\gamma }_{4}{\tilde{\Delta }}_{j}{\psi }_{j}^{n}}\right\} \eqno (7.7)$$

For the quantization of the Dirac field we require the equal time anticommutation relations
$$\lbrack{{\rm \Delta }_{j}{\psi }_{\alpha j}^{n},{\Delta }_{j'}{\psi }_{\beta j'}^{\dagger
n}}\rbrack_{+}={1 \over \varepsilon }{\delta }_{\alpha \rho }{\delta }_{jj'} \eqno (7.8)$$
with other anticommutations vanishing. If we plagg $H_n$ in the Heisenberg equations of motion,
we obtain the Dirac equation, from which the plane wave solutions and the Fourier
decomposition in terms of the creation and annihilation operators can be obtained (Lorente
1991).
 
\vskip 1.5truecm {\noindent \bf  8. CONSERVATION LAWS AND LORENTZ INVARIANCE} 

\medskip As in the continuous case we can make the connection between symmetries and
conservation laws in the language of generators. The condition for symmetry of the
Lagrangian under  space and time displacement and pure Lorentz transformation is that the
generators are constant of the motion (Yamamoto 1991). In the case of the Klein-Gordon fields the
generators of the (one step) space and time translations and Lorentz boost  can be taken as: 
 $$\eqalignno{ 
P&=-\sum\nolimits\limits_{j=0}^{N-1} {1 \over
2}\left\{{\left({{\tilde{\nabla}}_{j}{\tilde{\Delta }}_{j}{\pi
}_ {j}^{n}}\right)\left({{\tilde{\nabla }}_{j}{\Delta }_{j}{\phi
}_{j}^{n}}\right)+\left({{\tilde{\nabla }}_{j}{\Delta }_{j}{\phi
}_{j}^{n}}\right)\left({{\tilde{\nabla }}_{j}{\tilde{\Delta }}_{j}{\pi
}_{j}^{n}}\right)}\right\} &(8.1) \cr  
H&=\varepsilon \sum\nolimits\limits_{j=0}^{N-1} {1
\over 2}\left\{{{\left({{\tilde{\nabla }}_{j}{\tilde{\Delta }}_{j}{\pi
}_{j}^{n}}\right)}^{2}+{1 \over {\varepsilon }^{2}}{\left({{\tilde{\nabla }}_{j}{\Delta
}_{j}{\phi }_{j}^{n}}\right)}^{2}+{M}^{2}{\left({{\tilde{\nabla }}_{j}{\tilde{\Delta
}}_{j}{\phi }_{j}^{n}}\right)}^{2}}\right\} &(8.2)  \cr 
K&=\varepsilon
\sum\nolimits\limits_{j=0}^{N-1} {1 \over 2}\varepsilon j \left\{{{\left({{\tilde{\Delta
}}_{j}{\tilde{\nabla }}_{j}{\tilde{\Delta }}_{j}{\pi }_{j}^{n}}\right)}^{2}+{1 \over
{\varepsilon }^{2}}{\left({{\tilde{\Delta }}_{j}{\tilde{\nabla }}_{j}{{\Delta }_{j}\phi
}_{j}^{n}}\right)}^{2}+{M}^{2}{\left({{\tilde{\Delta }}_{j}{\tilde{\nabla
}}_{j}{\tilde{\Delta }}_{j}{\phi }_{j}^{n}}\right)}^{2}}\right\} \cr &\qquad -\varepsilon
\sum\nolimits\limits_{j=0}^{N-1} n\tau {1 \over 2}\left\{{\left({{\tilde{\Delta
}}_{j}{{\tilde{\nabla }}_{j}\phi }_{j}^{n}}\right)\left({{\tilde{\Delta }}_{j}{{\nabla
}_{j}\phi }_{j}^{n}}\right)+\left({{\tilde{\Delta }}_{j}{{\nabla }_{j}\phi
}_{j}^{n}}\right)\left({{\tilde{\Delta }}_{j}{{\tilde{\nabla }}_{j}\pi
}_{j}^{n}}\right)}\right\} &(8.3) \cr}$$ 

Using (6.13) and (6.14) it can be proved that these operators are constant of time  
$${\rm 1 \over \tau }{\Delta }_{n}P={1 \over \tau }{\Delta }_{n}H={1 \over \tau }{\Delta
}_{n}K=0 \eqno (8.4)$$ 

In order to check the Lorentz invariance of quantized field scheme on the lattice, one can
prove with the help of (6.16) and (6.17) that these operators satisfy the standard
commutation relations:  
$$[H,P]=0 \qquad , \qquad [K,H]=iP \qquad , \qquad [K,P]=iH \eqno (8.5)$$ 

For the Dirac quantum fields, the generators of the (one step) space and time
translations and Lorentz boost can be taken as  
$$\eqalignno{
P&=-i\sum\nolimits\limits_{j=0}^{N-1} \left(\tilde\Delta_j \psi_{\alpha j}^{\dagger
n}\right )\left(\Delta_j\psi_{\alpha j}^n \right ) &(8.6) \cr 
H&=\varepsilon\sum\nolimits\limits_{j=0}^{N-1}\tilde\Delta_j\psi_{\alpha j}^{\dagger
n}\left\{{\left(\gamma_4\gamma_1\right )}_{\alpha \beta }{1 \over \varepsilon
}\Delta_j\psi_{\beta j}^n + M{\left(\gamma _4\right )}_{\alpha \beta}
\tilde\Delta_j\psi_{\beta j}^n \right \} &(8.7) \cr  
M_{14}&=i\varepsilon
\sum\nolimits\limits_{j=0}^{N-1} \varepsilon j\left\{\tilde\Delta _j\tilde\nabla_j\psi_{\alpha
j}^{\dagger n}\left\{{\left(\gamma_4\gamma_1\right )}_{\alpha \beta}{1 \over \varepsilon
}\Delta_j\tilde\nabla_j\psi_{\beta j}^n + M\left(\gamma_4\right )_{\alpha
\beta}\tilde\Delta_j\tilde\nabla_j\psi_{\beta j}^n\right \}\right \} \cr 
&-i\varepsilon\sum\nolimits\limits_{j=0}^{N-1} \tau n\left\{\tilde\Delta_j\psi_{\alpha j}^{\dagger
n}\Delta_j\psi_{\alpha j}^n \right \}+\varepsilon
\sum\nolimits\limits_{j=0}^{N-1}\tilde\Delta\psi_{\alpha j}^\dagger{1 \over
2i}{\left(\gamma_1\gamma_4\right )}_{\alpha \beta }\tilde\Delta_j\psi_{\beta j}^n &(8.8) \cr
}$$  

Using (7.5) and (7.8) it can be proved that these operators are constant of time   
$${1\over \tau}\Delta_nP={1 \over \tau }\Delta_nH={1 \over \tau }\Delta_nM_{14}=0 \eqno
(8.9)$$  
and that they satisfy the standard commutation relations   
$$[H,P]=0 \quad , \quad [H,M_{14}]=P \quad , \quad [P,M_{14}]=H \eqno (8.10)$$ 

We can convince ourselves that $H$ and $P$ are the generators of the time and space
displacement by iteration of the Heisenberg equations of motion as in the continuous case. For the
operator $H$ we have $${ 1 \over i}\left[{{\tilde{\Delta }}_{n}\left({{\tilde{\nabla
}}_{j}{\tilde{\Delta }}_{j}{\phi }_{j}^{0}}\right), H}\right]={1 \over \tau }\left({{\tilde{\nabla
}}_{j}{\tilde{\Delta }}_{j}{\phi }_{j}^{0}}\right) \eqno (8.11) $$
where we have fixed the time index, say $n = 0.$

By iteration we have
$$\eqalignno{ 
\sum\nolimits\limits_{k=1}^{n} \left({\matrix{n\cr k\cr}}\right){\left({{\tau 
\over i}}\right)}^{k}{\left({{\tilde{\Delta }}_{n}}\right)}^{k} \underbrace
{\left[{{\tilde{\nabla}}_{j}{\tilde{\Delta }}_{j}{\phi }_{j}^{0},H}\right],\left.{H}\right],
\cdots,\left.{H}\right]}_{k\ {\rm times}} &=\sum\nolimits\limits_{k=1}^{n} \left({\matrix{n\cr k\cr}}\right){\Delta
}_{n}^{(k)}{\left({{\tilde{\nabla }}_{j}{\tilde{\Delta }}_{j}{\phi }_{j}^{0}}\right)}_{n=0} \cr
&={\tilde{\nabla }}_{j}{\tilde{\Delta }}_{j}{\phi }_{j}^{n}-{\tilde{\nabla }}_{j}{\tilde{\Delta
}}_{j}{\phi }_{j}^{0} &(8.12) \cr} $$
which can be taken as the ``Taylor expansion" on the lattice, namely,
$${\tilde{\rm \nabla }}_{j}{\tilde{\Delta }}_{j}{\phi }_{j}^{n}={\tilde{\nabla
}}_{j}{\tilde{\Delta }}_{j}{\phi }_{j}^{0}+\sum\nolimits\limits_{k=1}^{n} \left({\matrix{n\cr
k\cr}}\right){\Delta }_{n}^{(k)}{\left.{{\tilde{\nabla }}_{j}{\tilde{\Delta }}_{j}{\phi
}_{j}^{n}}\right|}_{n=0} \eqno (8.13)$$

In the limit $n\rightarrow \infty, \quad\tau\rightarrow 0,\quad
n\tau\rightarrow t$ \quad the expression (8.12) becomes
$$ \eqalignno{
\phi \left(x,0\right)+\sum\nolimits\limits_{k=1}^{\infty } {1 \over k!}{\left({{t
\over i}}\right)}^{k}\underbrace{\left[{\phi \left({x,0}\right),H}\right],\left.{H}\right],
\cdots,\left.{H}\right]}_{k\ {\rm times}} &=\phi\left({x,0}\right)+\sum\nolimits\limits_{k=1}^{\infty } {{t}^{k} \over k!}{\left.{{{\partial
}^{k} \over \partial {t}^{k}}\phi \left({x,t}\right)}\right|}_{t=0} \cr
&=\phi \left({x,t}\right) &(8.14) \cr}$$

But this expression is precisely the expansion of the continuous time traslations generated
by the operator $H$
$${ e}^{iHt}\phi \left({x,0}\right){e}^{-iHt}=\phi \left({x,t}\right) \eqno (8.15)$$

For the operator $P$ we have
$${1 \over i}\left[{{\tilde{\nabla }}_{j}{\tilde{\Delta }}_{j}{\phi
}_{0}^{n},P}\right]={1 \over \varepsilon }{\tilde{\nabla }}_{j}{\Delta }_{j}{\phi }_{0}^{n} \eqno
(8.16)$$ 
where we have fixed the space index, say $j=0$. By iteration we have
$$\displaylines{
\hskip 15 mm{\tilde{\rm \nabla }}_{j}{\tilde{\Delta }}_{j}{\phi}_{0}^{n}+\sum\nolimits\limits_{k=1}^{j}
\left({\matrix{j\cr k\cr}}\right){{\varepsilon }^{k} \over {i}^{k}}{\left({{\tilde{\Delta
}}_{j}}\right)}^{k-1}
\underbrace{\left[{{\tilde{\Delta}}_{j}{\phi}_{0}^{n},P}\right],\left.{P}\right],
\cdots,\left.{P}\right]}_{k {\rm times}} \hfill \cr \hfill ={\tilde{\nabla}}_{j}{\tilde{\Delta
}}_{j}{\phi }_{0}^{n}+\sum\nolimits\limits_{k=1}^{j} \left({\matrix{j\cr k\cr}}\right){{\Delta
}_{j}^{(k)}\left ({\tilde{\nabla }}_{j}{\phi}_{j}^{n}\right )}_{j=0} ={\tilde{\nabla
}}_{j}{\tilde{\Delta }}_{j}{\phi }_{j}^{n} \qquad (8.17)\cr} $$ which correspond to the $j$ step
space translation on the lattice. In the limit (8.17) becomes the continuous space translation
generated by the operator $P$.

\vskip 1.5truecm {\noindent \bf APPENDIX. THE EINSTEIN DE BROGLIE RELATION ON THE LA\-TTICE}

\medskip In order to make connection of our scheme with the Einstein-de Broglie relations
$E=\hbar \omega, \quad p=\hbar k$ we take the discrete plane waves solutions of (6.8). 
$${f}_{j}^{n}\left({k,\omega}\right)={\left({{1+{1 \over 2}i\varepsilon k \over
1-{1 \over 2}i\varepsilon k}}\right)}^{j}{\left({{1-{1 \over 2}i\tau \omega  \over 1+{1 \over
2}i\tau \omega }}\right)}^{n} \eqno (A.1)$$ 

Obviously, we have for the period $T$ and wave length $\lambda$
$$T=N\tau\quad , \quad\lambda=N\varepsilon \eqno (A.2)$$
and for the phase velocity 
$${ v}_{p}={\lambda  \over T}={\varepsilon  \over \tau } \eqno (A.3)$$

If we impose the boundary conditions   
$${ f}_{0}^{N}\left({k,\omega }\right)={f}_{N}^{n}\left({k,m}\right) \eqno (A.4)$$ 
the wave number and the angular frequency can be defined as
$${k}_{m}={2 \over \varepsilon }tg{\pi m \over N},\quad {\omega }_{m}=
{2 \over \tau }tg{\pi m \over N},\quad m=0,1,\cdots,N-1 \eqno (A.5)$$ 

Substituting the Einstein-de Broglie relations in the relativistic expresion $E^2-p^2$ $=M^2$ (we
use natural units $\hbar = c = 1$), we obtain 
$${ \omega }_{m}^{2}-{k}_{m}^{2}={\omega}_{m}^{2}\left({1-{{\tau }^{2} \over {\varepsilon
}^{2}}}\right)={\omega }_{m}^{2}\left({1-{1 \over {v}_{p}^{2}}}\right)={M}^{2} \eqno (A.6)$$

Since the phase velocity and group velocity satisfy $v_p v_g=1$, we have
finally 
$${\omega }_{m}^{2}={{M}^{2} \over 1-{v}_{g}^{2}} \eqno (A.7)$$ 
giving a discrete mass spectrum due to the lattice.

\vskip 1.5truecm  \noindent  {\bf ACKNOWLEDGMENT} 

\medskip We want to express our gratitude to Bruno Gruber for the invitation to
participate in this Symposium in honour of L. Biedenharn and for the oportunity to
present these ideas and discuss them during the symposium.
This work has been partially supported by Vicerrectorado de Investigaci\'on of Universidad
de Oviedo and by D.G.I.C.Y.T. (PS 89-0171). 

\vskip 1.5truecm  \noindent  {\bf REFERENCES} \medskip\hangindent=16pt \noindent Barut, A.O. and
Raczka, R. (1965). ``Classification of non-compact real Lie Groups and Groups containing the
Lorentz Group'', {\it Proc. Roy. Soc. London Series A, 287}, 519-548.

\smallskip\hangindent=32pt \noindent Beckers, J., Harnad, J., Perroud, M. and Winternitz, P.
(1978). ``Tensor field invariant under subgroups of the conformal group of space-time'', {\it J.
Math. Phys., 19}, 2126-2153.

\smallskip\hangindent=32pt \noindent Castell, L., Drieschner, M. and Weizsaecker, C.F. ed.
(1975-7-9-81-83-85). {\it Quantum Theory and the Structure of Space and Time}, Hanser, vol. 1-6.

\smallskip\hangindent=32pt \noindent Cayley, A. (1846). {\it Journal f\"ur reine und angewandte
Mathematik}, 32, 1 (1889). {\it Collected Mathematical Papers}, Cambridge, 117.

\smallskip\hangindent=32pt \noindent Earman, J. (1989). {\it World Enough and Space and Time, Relational Theories
of Space and Time}, Cambridge.

\smallskip\hangindent=32pt \noindent Gel`fan, I.M., Minlos, R.A. and Saphiro, Z. (1963). {\it Representations of
the Rotation and Lorentz Groups and their Applications}, Pergamon Press.

\smallskip\hangindent=32pt \noindent Gr\"unbaum, A. (1977). ``Absolute and Relational Theories of space and
Time'', in {\it Minnesota Studies in the Philosophy of Science} (J. Earman, C. Glymour, J. Stachel,
ed.). University of Minnesota Press, vol. VII.

\smallskip\hangindent=32pt \noindent Helgason, S. (1978). {\it Differential Geometry, Lie Groups and Symmetric
Spaces}. Academic Press, p. 518.

\smallskip\hangindent=32pt \noindent Jammer, M. (1969).{\it Concepts of Space and Time}. Cambridge. Harvard U.
Press.

\smallskip\hangindent=32pt \noindent Lorente, M. (1974). ``Cayley Parametrization of Semisimple Lie Groups and
its Application to Physical Laws in a (3+1)-Dimensional Cubic Lattice'', {\it Int. J. Theor. Phys. 11},
213-247.

\smallskip\hangindent=32pt \noindent Lorente, M. (1976). ``Basis for a Discrete Special Relativity'', {\it Int. J. Theor. Phys.
12}, 927.

\smallskip\hangindent=32pt \noindent Lorente, M. (1986a). ``Space-time Groups for the Lattice'', {\it Int. J.
Theor. Phys. 25}, 55-65.

\smallskip\hangindent=32pt \noindent Lorente, M. (1986b) ``A Causal Interpretation of the Structure of Space and
Time''. {\it Foundations of Physics}, H\"older Pichler-Tempsky, Viena.

\smallskip\hangindent=32pt \noindent Lorente, M. (1986c). ``Physical Models on Discrete Space and Time'', in
{\it Symmetries in Science II} (B. Gruber, R. Lenczweski, eds.), Plenum Press.

\smallskip\hangindent=32pt \noindent Lorente, M. (1987). ``The Method of Finite Differences for Some Operator
Field Equations'', {\it Lett. Math. Phys. 13}, 229-236.

\smallskip\hangindent=32pt \noindent Lorente, M. (1991). ``Lattice Fermions with Axial Anomaly and without
Species Doubling'', II Int. Wigner Symposium, Goslar.

\smallskip\hangindent=32pt \noindent Lorente, M. (1992). ``A Relativistic Invariant Scheme for the Quantum
Klein-Gordon and Dirac Fields on the Lattice''. XIX Int. Colloquium on Group Theoretical Methods in
Physics, Salamanca.

\smallskip\hangindent=32pt \noindent M\o ller, C. (1952). {\it The Theory of Relativity}, Oxford p. 42.

\smallskip\hangindent=32pt \noindent Naimark, M.A. (1964). {\it Linear Representation of the Lorentz Group},
Pergamon Press, p. 92.

\smallskip\hangindent=32pt \noindent Penrose R. (1971). ``Angular momentum: an approach to
combinatorial Space-time'' in {\it Quantum Theory and Beyond} (T. Bastin, ed.), Cambridge.

\smallskip\hangindent=32pt \noindent Schild, A. (1949). ``Discrete space-time and integral Lorentz
transformations'', {\it Canadian Journal of Mathematics, 1}, 29-47.

\smallskip\hangindent=32pt \noindent Wigner, E.P. (1959). {\it Group theory and Its Application to the Quantum
Mechanics of Atomic Spectra}, Academic Press, p. 160.

\smallskip\hangindent=32pt \noindent Yamamoto, H. (1985). {\it Phys. Rev. D32}, 2659. 

\smallskip\hangindent=32pt \noindent Yamamoto, H. (1991). ``Noether Theorem and Gauge Theory in the Field Theory
on Discrete Spacetime'', II Int. Wigner Symposium, Goslar.

\end